\documentclass[english,final,twocolumn]{revtex4}
\usepackage[T1]{fontenc}
\usepackage[latin9]{inputenc}
\usepackage{amsmath}
\usepackage{amssymb}
\usepackage{graphicx}
\usepackage{multirow}
\providecommand{\tabularnewline}{\\}
\makeatletter
\@ifundefined{textcolor}{}
{%
 \definecolor{BLACK}{gray}{0}
 \definecolor{WHITE}{gray}{1}
 \definecolor{RED}{rgb}{1,0,0}
 \definecolor{GREEN}{rgb}{0,1,0}
 \definecolor{BLUE}{rgb}{0,0,1}
 \definecolor{CYAN}{cmyk}{1,0,0,0}
 \definecolor{MAGENTA}{cmyk}{0,1,0,0}
 \definecolor{YELLOW}{cmyk}{0,0,1,0}
}

\makeatother

\usepackage{babel}
\begin{document}

\title{Two-Loop Effects in Low-Energy Electroweak Measurements}

\author{A. Aleksejevs}

\affiliation{Grenfell Campus of Memorial University, Corner Brook, Canada}

\author{S. Barkanova}

\affiliation{Acadia University, Wolfville, Canada }

\author{V. Zykunov}

\affiliation{Belarussian State University of Transport, Gomel, Belarus}

\begin{abstract}
We outline the recent results on the two-loop electroweak contributions
to the electron-electron scattering cross sections and asymmetries.
Although the two-loop corrections are strongly suppressed relative
to the one-loop corrections, they still contribute a few percent to
the polarization asymmetry, and even this small contribution cannot be ignored
at for ultra-precision experiments such as MOLLER planned at JLab.
The NNLO calculation techniques we developed for the electron-electron
scattering can be adapted for electron-proton processes, electron-positron
collisions, and other low-energy experiments involving leptons.
\end{abstract}
\maketitle

\section{Motivation}

There are three major ways to look for new physics beyond the Standard
Model (SM): the energy frontier (high-energy colliders), the intensity/precision
frontier (intense beams) and the cosmic frontier (underground experiments,
ground and space-based telescopes).
At the precision frontier, one of the most promising processes is
polarized electron-electron (M{\o}ller) scattering with parity violation,
potentially allowing the indirect detection of hypothetical new physics
particles coupling to the SM sector through the kinetic mixing.
The first measurement of the M{\o}ller scattering cross
section was done in 1932 \cite{M1932}, but only the recent improvements
in precision allowing to measure the parity-violating left-right asymmetry
made M{\o}ller scattering a candidate for the new-physics sector
detection.

The first observation of parity violation in M{\o}ller scattering
was made by the E-158 experiment at SLAC \cite{2}, which studied
M{\o}ller scattering of 45- to 48-GeV polarized electrons on the
unpolarized electrons in a hydrogen target. Its result at low $Q^{2}=0.026\mbox{\, GeV}^{2}$,
$A_{PV}=(1.31\pm0.14\ \mbox{(stat.)}\pm0.10\ \mbox{(syst.)})\times10^{-7}$
\cite{E158} allowed one of the most important parameters in the
Standard Model - the sine of the Weinberg angle (the weak mixing
angle) - to be determined with an accuracy of 0.5\% ($\sin^{2}\theta_{W}=0.2397\pm0.0010\pm0.0008$
in the ${\rm \overline{MS}}$ scheme).

A recently-completed JLab experiment measuring the electron-proton
scattering asymmetry, $Q_{weak}$ \cite{QWeak}, aims to determine
$\sin^{2}\theta_{W}$ with relative precision of 0.3\%.
The results of $Q_{weak}$ commissioning run, constituting about 4\%
 of the data collected \cite{QWeak2013}, give the left-right asymmetry
of $A_{PV}=-279\pm35\mbox{(stat)}\pm31\mbox{(syst)}\ \mbox{ppb}$,
the smallest and most precise asymmetry ever measured in e-p scattering,
and lead to the first determination of the weak charge of the proton,
$Q^p_W=0.064\pm0.012$, in agreement with the SM prediction of
$Q^p_W=0.0710\pm0.0007$.
From the theory aspect, the $Q_{weak}$ precision can be improved
by the better control of hadronic corrections and accounting for the
NNLO contributions to the electron line discussed in this work. With
that, and with the full set of data analyzed, $Q_{weak}$ has potential
to place tight constraints on the possible SM extensions.
Another PV e-p experiment, P2 proposed for the newly-funded MESA facility
at Mainz, aims to determine $Q^p_W$ even more precisely, to 2\%.

The next-generation experiment to study e-e scattering,
MOLLER (Measurement Of a Lepton Lepton Electroweak Reaction) \cite{Moller2014}, planned
at JLab following the 11 GeV upgrade, will offer a new level of sensitivity
and measure the PV asymmetry in the scattering of longitudinally polarized
electrons off an unpolarized target to a precision of 0.73 ppb. That
would allow a determination of the weak mixing angle with an uncertainty
of $\delta_{\sin^{2}\theta_{W}}(\overline{MS})=\pm0.00026\ \mbox{(stat.)}\pm0.00013\ \mbox{(syst.)}$
\cite{JLab12}, or about 0.1\%, an improvement of a factor of five
in fractional precision when compared with the E-158 measurement.
At such precision, any inconsistency with the Standard Model will
signal new physics, so the MOLLER experiment, building on the concept
of indirect probes, can provide access to physics at multi-TeV
scales.
The experiment will undoubtedly be more challenging than previous
parity-violating electron scattering experiments, but the MOLLER collaboration
has extensive experience in the similar experiments such as MIT Bates,
SLAC E158, JLab G0, HAPPEX, PREX and $Q_{weak}$.
The major advantage of the M{\o}ller scattering is that e-e scattering
asymmetry is much less affected by the uncertainties in the hadronic
corrections then e-p asymmetry, and the gamma-Z box radiative correction
to PV elastic e-p scattering calculated at 11 GeV in \cite{N42}
has an accuracy sufficient to keep the uncertainty from this background
within the limits of the MOLLER experiment.
The rest of the electroweak radiative corrections (EWC), although
extensive, can in principle be controlled at sub-1\% level, with the
SM predictions carried out with full treatment of one-loop radiative
corrections and at least leading two-loop corrections. 

It was repeatedly shown in the literature that even one-loop radiative corrections modify
the tree-level prediction for the asymmetry quite significantly (\cite{Czar1996},
\cite{Petr2003}, \cite{EM2005}, \cite{ABIZ-prd}, \cite{arx-2}),
so it is essential to have them under a very firm control. (Please
see \cite{KK2013} for a review of the low-energy measurements of
the weak mixing angle and additional references.)
In \cite{ABIZ-prd}, we found the total correction calculated specifically
for 11 GeV e-e scattering to be close to $\sim65\%$, with no significant
theoretical uncertainties. A much larger theoretical uncertainty in
the prediction for the asymmetry will come from the two-loop corrections,
so, for the new-generation precision measurements, predictions for
its scattering asymmetry must include not only a full treatment of
one-loop radiative corrections (NLO) but also leading two-loop corrections
(NNLO).

We approach the NNLO EWC in stages, by dividing the corrections to
the Born ($\sigma_{B}\sim|M_{0}|^{2}$) cross section into two classes:
the $Q$-part induced by quadratic one-loop amplitudes ($\sigma_{Q}\sim|M_{1}|^{2}$),
and the $T$-part corresponding to the interference of the Born and
two-loop diagrams ($\sigma_{T}\sim2\mbox{{\rm Re}}M_{2}M_{0}^{+}$).
The details of our calculations for the quadratic one-loop amplitudes,
the $Q$-part, are shown in \cite{Q-part}, where, following the
same approach we used for NLO EWC, we performed a tuned step-by-step
comparison between different calculation approaches verifying the
results obtained by a semi-automatic approach based on FeynArts, FormCalc,
LoopTools and Form with the results from the equations derived by
hand. As we found in \cite{Q-part}, for the MOLLER kinematic conditions,
the $Q$-part of the NNLO EWC can increase the asymmetry by up to
4\%, and depends quite significantly on the energy and scattering
angles.

In this paper, we discuss a set of contributions corresponding to
the interference of the Born and two-loop diagrams (the T-part), including
the gauge invariant set of boson self energies and vertices of two-loop
amplitude $M_{2}$, and discuss work still to be done in the future.

\section{Cross Section and Asymmetry}

The cross section of polarized M{\o}ller scattering
with the Born kinematics:
\begin{equation}
e^-(k_1)+e^-(p_1) \rightarrow e^-(k_2)+e^-(p_2),
\label{0}
\end{equation}
can be expressed as:
\begin{align}
\sigma & =\frac{\pi^{3}}{2s}|M_{0}+M_{1}+M_{2}|^{2}\approx\label{01}\\
 & \frac{\pi^{3}}{2s}(M_{0}M_{0}^{+}+2{\rm Re}M_{1}M_{0}^{+}+M_{1}M_{1}^{+}+2{\rm Re}M_{2}M_{0}^{+}),\nonumber 
\end{align}
where
$\sigma \equiv {d\sigma}/{d \cos \theta}$ and 
$\theta$  is the scattering angle of the detected electron
with 4-momentum $k_2$ in the center-of-mass system of the initial electrons. The
 4-momenta of initial ($k_1$ and $p_1$) and final
($k_2$ and $p_2$) electrons generate a standard
set of Mandelstam variables:
\begin{equation}
s=(k_1+p_1)^2,\ t=(k_1-k_2)^2,\ u=(k_2-p_1)^2.
\label{stu}
\end{equation}
$M_0$ is the Born (${\cal O}(\alpha)$) amplitude shown in Fig.1. 
The amplitudes $M_1$ (Fig.2) and $M_2$ (Figs.4-8) correspond 
to 
one-loop (${\cal O}(\alpha^2)$) 
and two-loop (${\cal O}(\alpha^3)$) 
matrix elements, respectively.

The one-loop amplitude $M_1$ consists of the boson self-energy (BSE) (Fig.2a),
vertex (Ver) (Fig.2b,c) and box diagrams (Fig.2d,e).
We use the on-shell renormalization scheme from \cite{BSH86, Denner},
so there are no contributions from the electron self-energies.

We present the one-loop complex amplitude as the sum of IR and IR-finite parts $M_1 = M_1^\lambda + M_1^f$. The IR-finite part $M_1^f$ can be found in \cite{Q-part} and for the IR part we have:
\begin{equation}
M_1^\lambda = \frac{\alpha}{2\pi}  {\delta_1^{\lambda}} M_0,\
\delta_1^{\lambda} = 4 B \log\frac{\lambda}{\sqrt{s}},
\label{mmm1}
\end{equation}
where
$\lambda$ is the photon mass and 
the complex value $B$ can be presented in the following form (see, for example, \cite{KuFa}):
\begin{equation}
B= \log\frac{tu}{m^2s}-1 - i\pi.
\end{equation}
Analogously, the two-loop amplitude is the sum $M_2 = M_2^\lambda + M_2^f$, where
\begin{eqnarray}
M_2^\lambda = 
  \frac{\alpha}{2\pi}  {\delta_1^{\lambda}}  M_1^f
+ \frac{1}{8}  \Bigl( \frac{\alpha}{\pi} 
 \Bigr)^2 \bigl( 
{\delta_1^{\lambda}} \bigr)^2 M_0.
\label{mmm2} 
\end{eqnarray}
Note that the structure of first term in (\ref{mmm2}) is the same as in (\ref{mmm1}) in terms 
of the soft photon factorization.

To cancel the infrared divergences, we split
the differential
cross sections into $\lambda$-dependent
(IRD-terms) and $\lambda$-independent (infrared-finite) parts:

\begin{eqnarray}
\sigma_1 = \sigma^{\lambda}_1 + \sigma^{f}_1,\ \
\sigma^V_{Q,T} =  \sigma^{\lambda}_{Q,T} + \sigma^{f}_{Q,T}.
\end{eqnarray}

The one-loop cross section is already carefully evaluated with full control
of the uncertainties in \cite{ABIZ-prd}.
The simplest form for IRD-terms are:


\begin{align}
 & \sigma_{1}^{\lambda}=\frac{\alpha}{\pi}{\rm Re}\bigl(\delta_{1}^{\lambda}\bigr)\sigma_{0},\nonumber \\
 & \sigma_{Q}^{\lambda}=\Bigl(\frac{\alpha}{2\pi}\Bigr)^{2}\Bigl[|\delta_{1}^{\lambda}|^2 + 2{\rm Re}\bigl(\delta_{1}^{f}\delta_{1}^{\lambda*}\bigr)\Bigr]\sigma^{0},\\
 & \sigma_{T}^{\lambda}=\Bigl(\frac{\alpha}{2\pi}\Bigr)^{2}{\rm Re}\Bigl[(\delta_{1}^{\lambda}){}^{2}+2\bigl(\delta_{1}^{f}\delta_{1}^{\lambda}\bigr)\Bigr]\sigma^{0}.\nonumber 
\end{align}

The imaginary part of the total cross section could be removed in the sum $Q$- 
and $T$-parts due to following properties: 
$  |\delta_1^{\lambda}|^2 + {\rm Re} {\bigl( \delta_1^{\lambda} \bigr)}^2
  =  2 {\bigl( {\rm Re} \ \delta_1^{\lambda} \bigr)}^2$,
and 
$ {\rm Re} \bigl( \delta_1^{f} {\delta_1^{\lambda}}^* \bigr) 
+  {\rm Re} \bigl( \delta_1^{f} {\delta_1^{\lambda}} \bigr) = 
 {\rm Re} (\delta_1^{f}) \ {\rm Re}  ({\delta_1^{\lambda}})$.
Thus, in
the following sections we can ignore the imaginary part, i. e. 
$ \delta_1^{\lambda} \rightarrow {\rm Re} \delta_1^{\lambda} $ and
$ B \rightarrow {\rm Re} B $.

\section{Bremsstrahlung for NLO and NNLO}

Bremsstrahlung for both NLO (Fig.3a) and NNLO (Fig.3b,c) is needed to cancel the infrared divergences in the one-loop and two-loop amplitudes, correspondingly. (Radiation from only one lepton line is shown in Fig.3, but all four lepton lines are accounted for in our calculations, of course.)
To evaluate the cross section induced by the emission of one soft photon with energy less then $\omega$,
we follow the methods of  \cite{HooftVeltman} (see also	\cite{KT1}).
Then, this cross section can be expressed as:
$ \sigma^{\gamma}=  \sigma^{\gamma}_1 + \sigma^{\gamma}_2$,
where $\sigma^{\gamma}_{1,2}$ have the similar factorized  structure
based on the factorization of the soft-photon bremsstrahlung:

\begin{equation}
 \sigma^{\gamma}_{1,2}= \frac{\alpha}{\pi} \bigl[ -\delta_1^{\lambda} +R_1 \bigr] \sigma_{0,1},
\label{SB}
\end{equation}

where
\begin{equation}
R_1=-4B \log\frac{\sqrt{s}}{2\omega} - \Bigl( \log\frac{s}{m^2} - 1\Bigr)^2 +1-\frac{\pi^2}{3} +\log^2\frac{u}{t}.
\end{equation}
The first part of the soft-photon cross section,  $\sigma^{\gamma}_1$,  cancels the IRD at the one-loop order,
while the second part, $\sigma^{\gamma}_2$, cancels the IRD at the two-loop order,
with half of $\sigma^{\gamma}_2$ going to the cancellation of the IRD in the $Q$-part and the
other half going to treat IRD in the $T$-part.
At last, the cross section induced by the emission of two soft photons with a total energy less then $\omega$
is calculated in \cite{Q-part} as:
\begin{equation}
\sigma^{\gamma\gamma}=
\frac{1}{2}
{\Bigl( \frac{\alpha}{\pi} \Bigr)}^2
\bigl( \bigl( -\delta_1^{\lambda} + R_1 \bigr)^2 - R_2 \bigr)
\sigma_0,
\end{equation}
where $\frac{1}{2}$ is a statistical factor 
and $R_2 = \frac{8}{3}\pi^2 B^2$.

Bringing all terms together, we arrive at the result that is free from infrared divergences. For one loop, the logarithms will cancel out:

\begin{equation}
\sigma_{{\rm NLO}}=\sigma_{1}+\sigma_{1}^{\gamma}=\frac{\alpha}{\pi}[R_{1}+\delta_{1}^{f}]\sigma^{0}\label{1IR}.
\end{equation}
For the second loop, the cancellation proceeds in a more involved way, that is
\begin{align}
\sigma_{{\rm NNLO}} & =\sigma_{Q}^{V}+\sigma_{T}^{V}+\sigma_{2}^{\gamma}+\sigma^{\gamma\gamma}\nonumber \\
 & =\Bigl(\frac{\alpha}{\pi}\Bigr)^{2}[R_{1}\delta_{1}^{f}+\frac{1}{2}R_{1}^{2}-\frac{1}{2}R_{2}+\delta_{Q}^{f}+\delta_{T}^{f}]\sigma^{0}\nonumber \\
 & =\sigma_{O}^{f}+\sigma_{B}^{f}+\sigma_{Q}^{f}+\sigma_{T}^{f},\label{2IR}
\end{align}
where 
\begin{eqnarray}
\sigma_{O}^{f}=\frac{\alpha}{\pi}R_{1}\sigma_{{\rm NLO}},\ \sigma_{B}^{f}=-\frac{1}{2}\Bigl(\frac{\alpha}{\pi}\Bigr)^{2}(R_{1}^{2}+R_{2})\sigma^{0}.\label{O-i-B}
\end{eqnarray}

\section{Numerical Results}

For the numerical calculations at the central kinematic point of MOLLER
($E_{lab}$=11 GeV, $\theta =\pi/2$)
we use
$\alpha$,\ $m_W$,  $m_Z$ and lepton masses as input parameters in accordance with \cite{PDG08}.
The effective quark masses which we use for the vector boson self-energy loop contributions are extracted from shifts in the 
fine structure constant due to hadronic
vacuum polarization $\Delta \alpha_{had}^{(5)}(m_Z^2)$=0.02757 \cite{jeger}.
For the mass of the Higgs boson, we take $m_H=125\ \mbox{GeV}$ 
and for the maximum soft photon energy we use $\omega = 0.05\sqrt{s}$, 
according to   \cite{ABIZ-prd} and \cite{5-DePo}.

Let us define the relative corrections to the Born cross section due to a 
specific type of contributions (labeled by $C$) as
$$\delta^{C} = (\sigma^{C}-\sigma^0)/\sigma^0,\ \ C=\mbox{NLO}, O, B, Q, T, \mbox{NNLO}.$$
In the text below the term "$T$-part" corresponds to the contributions of a gauge invariant set of the BSE
and vertices only. The parity-violating asymmetry is defined in a traditional way:
\begin{equation}
A_{LR} =\frac{\sigma_{LL}-\sigma_{RR}}
      {\sigma_{LL}+2\sigma_{LR}+\sigma_{RR}},
\label{A}
\end{equation}
and the relative corrections to the Born asymmetry  due to  $C$-contribution
are defined as
$$\delta^{C}_A = (A_{LR}^{C}-A_{LR}^0)/A_{LR}^0.$$
In general, corrections from different diagrams are not additive. Their total contribution is given by

\begin{equation}
\displaystyle{\delta^{\Sigma C_i}_A=\frac{\Sigma (1+\delta^{C_i})\delta^{C_i}_A}{1+\Sigma \delta^{C_i}}},
\label{AA}
\end{equation}
where summation is performed over the index $i$. There are reasons to believe that the correction $\delta^{NNLO}$ is small, but we can not say the same about  $\delta^{NNLO}_A$.

In the table below we bring together all contributions derived for relative corrections to the unpolarized cross section to the asymmetry, including contributions that stem from the gauge-invariant set of two-loop vertex and boson-self-energy diagrams. The three dots mean the contribution from the line above, so we progressively add new contributions as we have them ready. So far, as one can see from the table, the $Q$-part induced by quadratic one-loop amplitudes ($ \sim M_1M_1^+$),
and the contributions to $T$-part corresponding to the interference of the Born and 
two-loop diagrams ($ \sim 2 \mbox{\rm Re} M_{2} M_{0}^+$) considered here shift the result in the same direction. Under the kinematic conditions of the MOLLER experiment, the asymmetry that takes into account the concerted effect of one- and two-loop contributions decreases by about 62.7\%. For comparison, the one-loop contribution yield a value of -69.5\%. Clearly, there is still a lot to be done, and no definite conclusion can be made until all contributions are accounted for, but it looks like if no major cancellations are introduced by the remaining two-loop contributions, the NNLO effect on the PV asymmetry may be more significant that previously believed. Thus, it is safe to say even now that the research program for the MOLLER experiment must include evaluation of the full set of two-loop corrections. Although our numerical calculations are done for the MOLLER the kinematic conditions, the analytics will be directly applicable for the collider experiments, so we assume the NLO and NNLO contributions will be affecting their cross section asymmetry significantly as well, and will need to be evaluated for these measurements in the future. 

\begin{center}
\begin{tabular}{|c|c|c|c|}
\hline 
Type of & \multirow{2}{*}{$\delta^{c}$} & \multirow{2}{*}{$\delta_{A}^{C}$} & \multirow{2}{*}{Published}\tabularnewline
contribution &  &  & \tabularnewline
\hline 
\hline 
NLO & -0.1145 & -0.6953 & \cite{ABIZ-prd}\tabularnewline
\hline 
...+(O+B)/2+Q & -0.1125 & -0.6536 & \cite{Q-part}\tabularnewline
\hline 
...+(O+B)/2+BBSE & \multirow{2}{*}{-0.1201} & \multirow{2}{*}{-0.6420} &\multirow{2}{*}{\cite{YadFiz2013}} \tabularnewline
+VVer+VerBSE &  &  & \tabularnewline
\hline 
...+ double boxes & -0.1201 & -0.6534 & \cite{EurPh2012}\tabularnewline
\hline 
...+NNLO QED & -0.1152 & -0.6500 & \tabularnewline
\hline 
...+SE and  & \multirow{2}{*}{-0.1152} & \multirow{2}{*}{-0.6503} & \multirow{2}{*}{}\tabularnewline
Ver in boxes &  &  & \tabularnewline
\hline 
...+NNLO EW Ver & \multicolumn{3}{c|}{under way}\tabularnewline
\hline 
\end{tabular}
\par\end{center}

\begin{figure*}
\centering{}\includegraphics[scale=0.75]{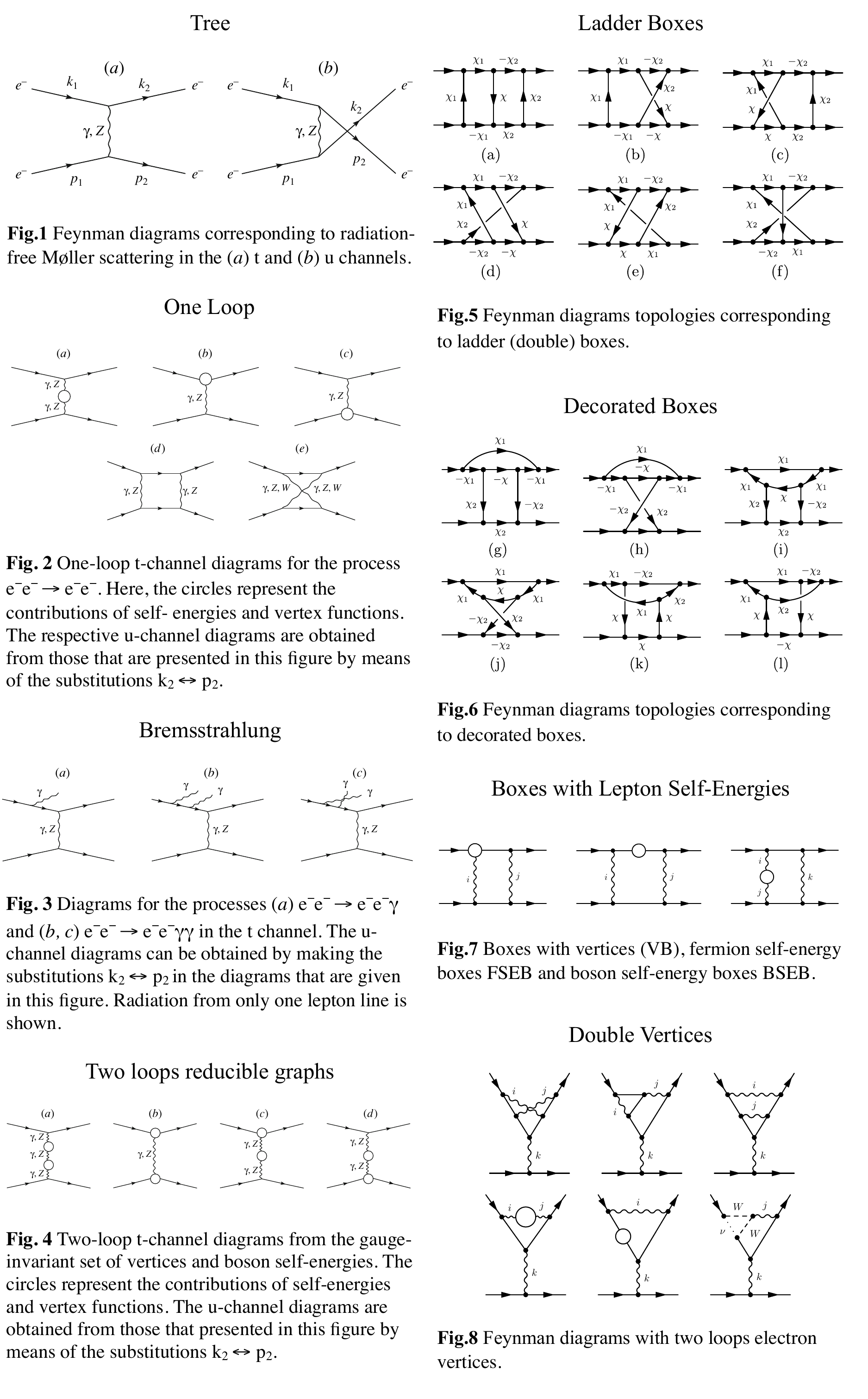}
\end{figure*}

\section{Conclusion}

As one can see from our numerical data, at the MOLLER kinematic conditions, the part of
the NNLO EWC we considered in this work can increase the asymmetry by up to $\sim$ 7\%. 
The $Q$- and $T$-parts do not cancel each other but, on the contrary, they are adding up to increase the physical PV effect.
Clearly, the large size of the investigated parts demands a detailed and consistent consideration of
the rest of the $T$-part, which will be the next task of our group. Since the problem of EWC for the
M{\o}ller scattering asymmetry is rather involved, a tuned step-by-step comparison between different
calculation approaches is essential. To make sure that our calculations are error-free, we control our results by comparing the data obtained from the equations
derived by hand with the numerical data obtained with a semi-automatic approach based on FeynArts,
FormCalc, LoopTools and Form. These base languages have already been successfully employed
in similar projects \cite{ABIZ-prd} and \cite{arx-2}, 
so we are highly confident in their reliability. 

In the future, we plan to address the remaining two-loop electroweak corrections to match the planned precision of the MOLLER experiment and the possible future experiments at ILC.
Clearly, for the electroweak low-energy experiments briefly outlined in this paper and for other potential future measurements, 
it is absolutely essential for an excellent control of one-loop and good understanding of two-loop effects \cite{KK2013}. 

In this paper, we outlined motivation and summarized some of our work on the two-loop electroweak radiative corrections involving the SM particles. 
Even if the LHC continues to agree with the Standard Model up to 14 TeV, the MOLLER experiment will continue to look for new physics scenarios that could escape LHC detection, 
like various hidden weak scale scenarios. If the LHC does observe an anomaly, then MOLLER will have enough sensitivity to provide sufficient constraints to distinguish between
 the possible new physics scenarios (new massive or super-massive $Z_0$ bosons, for example). To have that kind of sensitivity, the MOLLER aims to measure the PV asymmetry
 predicted within SM to be about 33 ppb with an overall precision of 0.7 ppb. The advantage of trying to access new physics via such low-energy e-e scattering asymmetry
 is that a purely leptonic PV asymmetry is one of the few observables whose theoretical uncertainties are well under control. There is no significant contribution from
 the hadronic sector, the SM Higgs mass, one of our input parameters, is known well enough for our needs, and the full set of NLO (one-loop) electroweak radiative
 corrections, although large, is now known to better than 0.1\%. 
Just a decade ago, such precision would not be feasible. Now, with the recent development in computer algebra and increased accessibility of computing facilities,
 we can aim to further improve the SM prediction for PV asymmetry by calculating the radiative corrections at the NNLO (two-loop) level.
 Since the EWC corrections depend quite significantly on the energy and scattering angles, they would need to be evaluated for each experiment specifically.
 For example, at the MOLLER kinematic conditions, the part of EWC induced by quadratic one-loop amplitudes ($ \sim M_1M_1^+$) will increase the asymmetry up to ~4\%, 
but increases dramatically in the higher-energy region \cite {Q-part}. This by itself is not a problem, since the $Q$-part is now know. However, we still far from making 
the final conclusion on behavior of the $T$-part corresponding to the interference of the Born and two-loop diagrams ($ \sim 2 \mbox{\rm Re} M_{2} M_{0}^+$).
So far, dominant two-loops contributions to the PV asymmetry are at the order of 1\% and they
are coming from (Ver + BSE)$^2$ and double boxes. 
As far as we know at the moment, the new-physics particles are not going to contribute significantly enough at two loops to warrant full-scale calculations, but they may contribute quite noticeably at the one-loop level, depending on the SM extension employed.

\section{ACKNOWLEDGMENTS}

The authors gratefully acknowledge Yu. Bystritskiy and \framebox{E. Kuraev} for their help with this project, and the the Joint Institute for Nuclear Research for hospitality in 2013. This work is supported by the
Natural Sciences and Engineering Research Council of Canada and Belarus scientific program "Convergence". AA and SB  thank JLab Theory Group for 
hospitality during their stay in 2014.

\end{document}